\documentclass[twoside,twocolumn,9pt]{article}
\usepackage{extsizes}
\usepackage[super,sort&compress,comma]{natbib} 
\usepackage[left=1.5cm, right=1.5cm, top=1.785cm, bottom=2.0cm]{geometry}
\usepackage{balance}
\usepackage{sectsty}
\allsectionsfont{\scshape}

\usepackage{graphicx} 
\usepackage{lastpage}
\usepackage{float}
\usepackage{fancyhdr}
\usepackage{fnpos}
\usepackage[english]{babel}
\usepackage{array}
\usepackage[T1]{fontenc}
\usepackage[usenames,dvipsnames]{xcolor}
\usepackage{setspace}
\usepackage{hyperref}
\usepackage[squaren]{SIunits}
\usepackage{amssymb}

\usepackage{lipsum}  
\usepackage{upgreek}
\usepackage{xspace}
\usepackage{placeins}
\usepackage{amsmath,amssymb}
\usepackage{color}  

\newcommand{\hmin}{\ensuremath{h_\mathrm{min}}\xspace}
\newcommand{\ho}{\ensuremath{h_\mathrm{0}}\xspace}
\newcommand{\eps}{\ensuremath{\dot{\varepsilon}}\xspace}

\newcommand{\etas}{\ensuremath{\eta_\mathrm{s}}\xspace}
\newcommand{\lmr}{\ensuremath{\lambda_\mathrm{R}}\xspace}
\newcommand{\lma}{\ensuremath{\lambda_\mathrm{a}}\xspace} 
\newcommand{\gmd}{\ensuremath{\dot{\gamma}}\xspace}
\newcommand{\tc}{\ensuremath{t_\mathrm{c}}\xspace}

\newcommand{\com}[1]{{{\color{black}\noindent#1}}} 

\newcommand{\finalCom}[1]{\textcolor{black}{#1}}

\begin{document}

\twocolumn[
  \begin{@twocolumnfalse}
\vspace{3cm}

\begin{center}

    \noindent\huge{\textbf{\textsc{Pinch-off of bubbles in a polymer solution}}} \\
    \vspace{1cm}

    \noindent\large{Sreeram Rajesh,\textit{$^{a}$} Sumukh S Peddada,\textit{$^{a}$} Virgile Thi\'evenaz,\textit{$^{a}$} and Alban Sauret\textit{$^{a}$}}$^{\ast}$ \\

    \vspace{5mm}
    \noindent\large{\today} \\

    \vspace{1cm}
    \textbf{\textsc{Abstract}}
    \vspace{2mm}

\end{center}

\noindent\normalsize{The formation of gas bubbles in a liquid occurs in various engineering processes, such as during foam generation or agitation and mixing in bubbly flows. A challenge in describing the initial formation of a gas bubble is due to the singular behavior at pinch-off. Past experiments in Newtonian fluids have shown that the minimum neck radius follows a power-law evolution shortly before the break-up. The exponent of the power-law depends on the viscosity of the surrounding Newtonian liquid, and ranges from 0.5 for low viscosity to 1 for large viscosity. However, bubble formation in a viscoelastic polymer solution remains unclear, and in particular, if the evolution is still captured by a power-law and how the exponent varies with the polymer concentration. In this study, we use high-speed imaging to analyze the bubble pinch-off in solutions of polymers. We characterize the time evolution of the neck radius when varying the concentration and thus the characteristic relaxation time and describe the influence of viscoelasticity on the bubble pinch-off. Our results reveal that the presence of polymers does not influence the thinning until the latter stages, when their presence in sufficient concentration delays the pinch-off.} \\

 \end{@twocolumnfalse} \vspace{0.6cm}

  ]

\makeatletter
\renewcommand*{\@makefnmark}{}
\footnotetext{\textit{$^{a}$~Department of Mechanical Engineering, University of California, Santa Barbara, California 93106, USA}}
\footnotetext{\textit{$^{\ast}$ asauret@ucsb.edu}}
\makeatother


\section{Introduction}

\begin{figure}[ht]
  \begin{center}
    \includegraphics[width=0.84\linewidth]{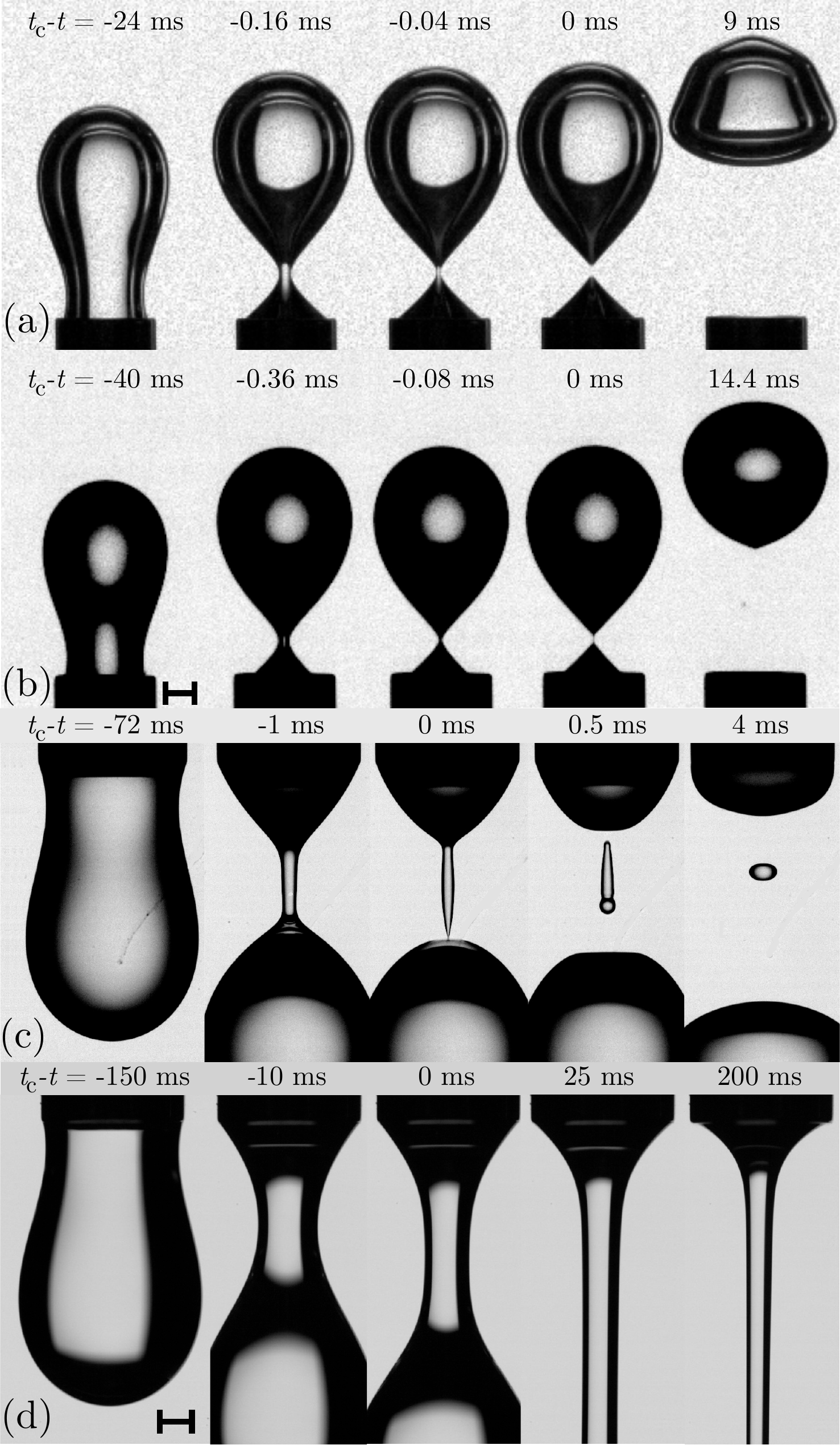}
  \end{center}
  \caption{Examples of a gas bubble pinch-off in a quiescent liquid of (a) 75/25\% by weight water/glycerol, and (b) a polymer solution (0.5\% mass concentration of 4000K PEO in 75/25\% by weight water/glycerol). Examples of pinch-off in air of a liquid droplet of (c) 75/25\% by weight water/glycerol ($\etas = 2.14\, \milli \pascal \second$), and (d) a polymer solution (0.5\% mass concentration of 4000K PEO in a 75/25\% by weight water/glycerol). Scale bars are 1 mm. }
  \label{fgr:figure1}
\end{figure}

Bubbles are encountered in a wide range of situations such as industrial processes,\cite{Crawford2017} biological systems,\cite{VanLiew1969} and in geological studies.\cite{parmigiani_bubble_2016} In the medical industry, bubbles have been used as contrast agents for ultrasound scans.\cite{Blomley2001} Compound bubbles formed by water contaminated with harmful substances due to the possible aerosolization are of concern to our health.\cite{ji2021oila,ji2021oil,ji_oil-coated_2021} Preventing cavitation bubbles is of crucial importance in biological networks in plants,\cite{Brodribb_PNAS_2016} or in the design of underwater turbines and propellers.\cite{Plesset1949} The formation of bubbles are also of particular interest in microfluidic devices,\cite{vanHoeve2011} microcapillary tubes,\cite{pahlavan_restoring_2019} as well as in turbulent flows.\cite{ruth_bubble_2019} In many applications, the liquids involved have a more complex rheology. For instance, the presence of cells, particles, or polymeric substances dispersed in the liquid modifies their rheology.\cite{Martinez2014,Patteson2015,Kamdar2022, Furbank2004,ThievenazBidisperse2021,ThievenazSuspensions,Narvaez2021} The modified rheology of polymer solutions has been exploited to achieve drag reduction in flow,\cite{sreenivasan_white_2000} and suppress the effect of cavitation.\cite{brujan1996dynamics} It is known that the presence of polymers modifies the pinch-off of a liquid droplet in air.\cite{Amarouchene2001,Tirtaatmadja2006} The inverse problem of pinch-off of an air bubble in a polymer solution has received less attention. 

The pinch-off of a gas bubble in a Newtonian liquid, shown in Fig. \ref{fgr:figure1}(a), is usually quantified through the time evolution of the minimum thickness $\hmin$ of the neck that connects the bubble with the nozzle. The moment the bubble detaches from the nozzle in a Newtonian fluid is a singularity in time $\tc$. For a short duration before the pinch-off, the minimum thickness follows a power law, $\hmin (t) \propto (\tc-t)^\alpha$. Burton \textit{et al.}\cite{burton_scaling_2005} have shown that the exponent $\alpha$ is a function of the external liquid viscosity for a Newtonian liquid. For inviscid liquids ($\eta \lesssim 10\,\milli\pascal\,\second$), the exponent is around $0.5$ and for viscous liquids ($\eta \gtrsim 100\,\milli\pascal\,\second$), it is approximately $1$. For liquids of intermediate viscosities, the exponent was reported in the range $0.5$ to $1$. A follow-up work by Thoroddsen \textit{et al.}\cite{thoroddsen_experiments_2007} at a higher spatial and temporal resolution reported an exponent of $0.57$ for inviscid Newtonian liquids. This slightly larger exponent was theoretically derived by Eggers \textit{et al.} as $\alpha = 1/2 + 1/[4\sqrt{-\ln(\tc-t)}]$ for the collapse of an axisymmetric cavity.\cite{eggers_theory_2007} \com{The exponents are hence non-universal, and depends on the initial condition of the system and the experimental resolution.} In this study, we do not focus on the influence of the initial condition. Instead, we use a spatial and temporal resolution similar to Burton \textsl{et al.}, which is sufficient to examine the moment leading up to the bubble pinch-off in polymer solutions. An example of thinning and pinch-off of a bubble in a polymer solution is shown in Fig. \ref{fgr:figure1}(b).

In the opposite configuration, \textit{i.e.}, when a drop of inviscid liquid thins in air, the minimum thickness is described by the power-law $\hmin(t) \propto (\tc-t)^{2/3}$.\cite{Kellert1983} Here, the time $\tc$ describes the moment the drop separates from the liquid attached to the nozzle, as shown in Fig. \ref{fgr:figure1}(c). Until a time $\tc$, the thinning of polymer solution, shown in Fig. \ref{fgr:figure1}(d), is also captured by a power law.\cite{Rajesh2022} At $\tc$, while the drop breaks off for a Newtonian liquid, adding even a small amount of polymer results in a transition to a viscoelastic thinning.\cite{Amarouchene2001} At the transition, the polymers, which are initially coiled, begins to unwind and interact with the flow.\cite{DeGennes1974,Rajesh2022} A macroscopic manifestation of this interaction is the formation of a long and slender liquid thread, which persists for a long time. \com{A further consequence of polymer uncoiling is the increase in the flow resistance. When the polymers are coiled, their hydrodynamic interaction with the solvent are minimized. As the polymer unwinds, more regions of the chain are exposed to the solvent. This results in an increase in the viscosity of the solution, \textit{i.e.}, an extensional thickening effect.\cite{Amarouchene2001,dinic2019macromolecular, dinic2015extensional}} In this regime, the slender filament thins exponentially following $\hmin(t) \propto {\rm exp}[-t/(3\,\lmr)]$, where $\lmr$ is the longest relaxation time of the polymer.\cite{Anna2001} The exponential thinning is followed by either a beads-on-a-string (BOAS) instability,\cite{clasen2006beads} or a blistering instability.\cite{Deblais2018}

While the pinch-off of bubbles in Newtonian liquids are well characterized,\cite{pahlavan_restoring_2019, Bergmann2009} there is a dearth in the literature on bubble pinch-off in viscoelastic liquids. At a larger scale, notable differences exist for bubbles in viscoelastic liquids, such as the negative wake reported by Hassager.\cite{hassager1979} The problem of discontinuity of the terminal velocity with respect to the bubble volume in polymer solutions has also been of long-standing interest.\cite{Leal1971,Bothe2022} The scales associated with the pinch-off are more challenging to capture and require a spatio-temporal resolution of a few microns over a few microseconds. A recent work by Jiang \textit{et al.}\cite{jiang_bubble_2017} has reported the existence of two distinct regimes during the final stages of the bubble pinch-off in polymer solutions of higher concentrations. At lower polymer concentrations, the thinning is a power-law, with the exponent $0.5 < \alpha < 1$. This value is larger than the thinning exponent for the solvent, which is $\alpha = 0.5$. At higher concentrations, there is no clear description for the thinning. 

\com{Yet, a recent study on the coalescence of two drops of liquid}, which is another example of singular behavior,\cite{paulsen2014coalescence} has shown that the minimum thickness of the coalescing polymer solutions is also described by a power-law and with the same exponent as the solvent.\cite{dekker_2022} The polymer solutions used in the coalescence study exhibit weak shear thinning and strong elasticity, \textit{i.e.} a sufficiently large relaxation time. In contrast, the solutions used by Jiang \textsl{et al.} for their pinch-off experiments have strong shear thinning and elasticity for all the concentrations studied, with relaxation times of the order of a few seconds. In particular, the change in exponent observed by Jiang \textit{et al.}\cite{jiang_bubble_2017} could be induced by the strong shear thinning of their solutions or the elastic effects, or by both.  

\begin{figure}[t]
  \begin{center}
    \includegraphics[width=1\linewidth]{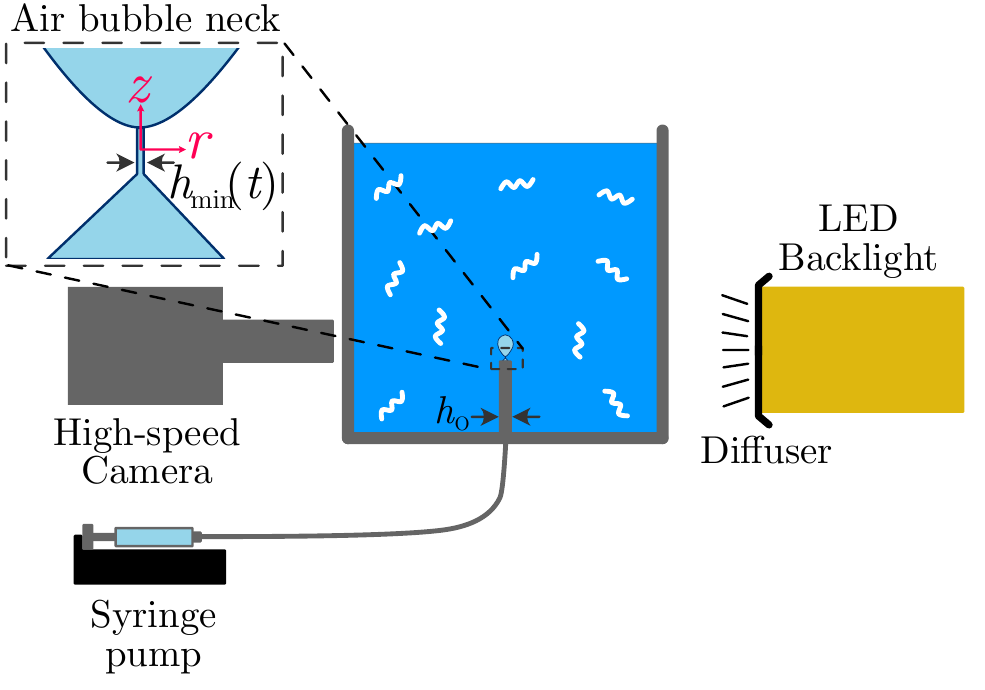}
  \end{center}
  \caption{Schematic of the experimental setup. A needle of internal diameter $\ho =2.31\,{\rm mm}$ is set at the bottom of a tank filled with the liquid and connected to a syringe filled with air. A syringe pump is used to generate a bubble from the nozzle. The length \hmin corresponds to the minimal thickness at the neck of the bubble. \com{The $z$-axis corresponds to the axis of symmetry at the neck.} The system is backlit with a LED (Light Emitting Diode) and a diffuser; the dynamics is recorded with a high-speed camera.} 
  \label{fgr:figure2}
\end{figure}

\begin{figure*}[ht]
  \begin{center}
    \includegraphics[width=1\linewidth]{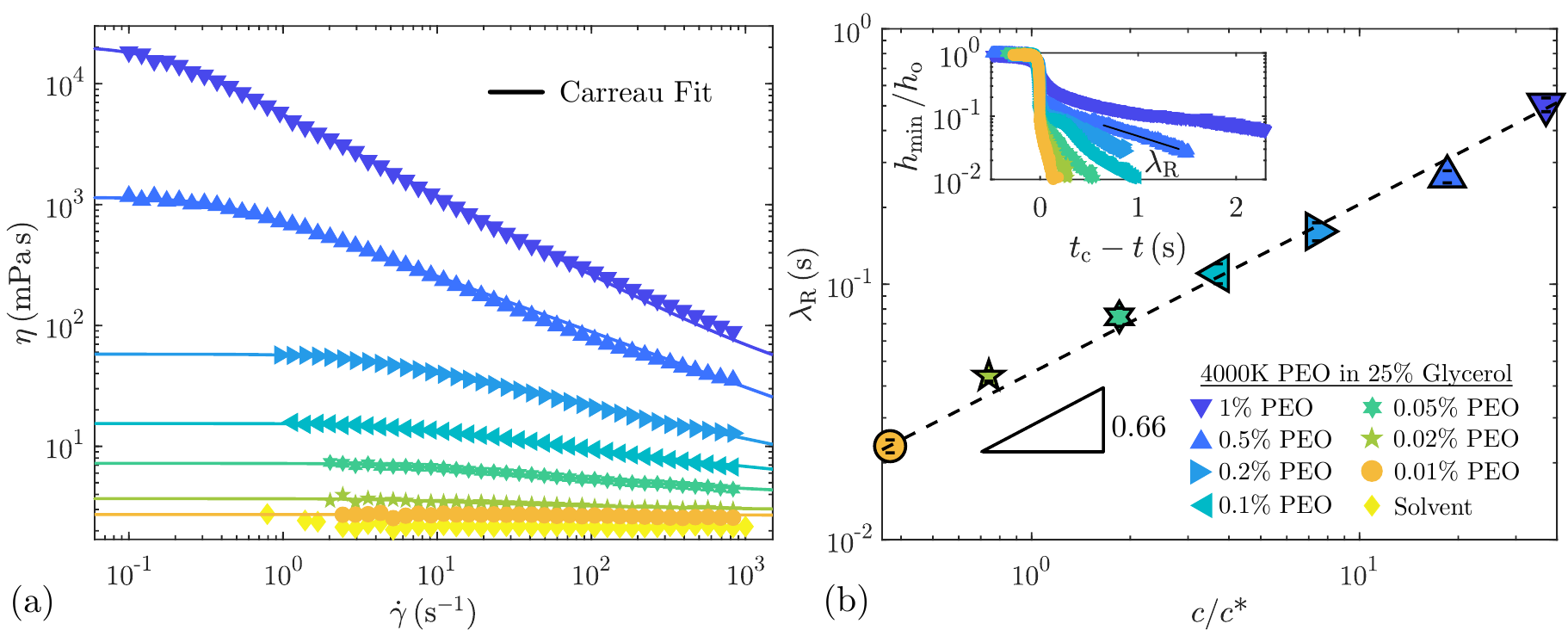}
  \end{center}
  \caption{Physical properties of the 4000K PEO polymer solutions with mass concentration between $c = 0.01\%$ and $c = 1\%$ prepared in a 75/25\% by weight water/glycerol solvent. (a) Shear viscosity $\eta$ of the solutions as a function of the shear rate $\dot{\gamma}$. (b) Relaxation time $\lmr$ of the solutions measured from the droplet thinning experiments. The increase in the relaxation times with the concentration follows a power-law with an exponent $0.66$. Inset: Thinning dynamics of PEO solutions. We extract the rescaled minimum thickness $\hmin/\ho$ as a function of time. The relaxation time $\lmr$ is obtained from the slope when $t-\tc > 0$, \textit{i.e.}, after the transition to the viscoelastic regime.}
  \label{fgr:figure3}
\end{figure*}

\begin{table*}[ht]
\centering
\begin{tabular}{ |c|c|c|c|c|c|c|c| } 
\hline
c & $c/c^*$ & $\sigma$ & $\lmr$ & $\eta_{\rm 0}$  & $\eta_{\rm \infty}$  & $\lambda$ & $n_{\rm cf}$ \\
        &       &  [mN/m] & [s]  & [mPa s] &[mPa s]& [s]  &       \\
\hline
 0\%    & 0     & 72.7   &   -   & -       &  -    & -     & -    \\ 
 0.01\% & 0.37  & 62.3  & 0.023 & 2.7    & 1     & 0.041 & 1    \\ 
 0.02\% & 0.74  & 63.8  & 0.043 & 3.7    & 2.89  & 0.23  & 0.7  \\
 0.05\% & 1.85  & 63.8  & 0.074 & 7.2    & 3.02  & 0.2   & 0.8  \\
 0.1\%  & 3.71  & 64.0     & 0.11  & 15.4   & 2.61  & 0.19  & 0.79 \\
 0.2\%  & 7.41  & 63.8  & 0.16  & 57.9   & 0.52  & 0.36  & 0.72 \\
 0.5\%  & 18.52 & 63.6   & 0.26  & 1150 & 0.1   & 2.63  & 0.54 \\
 1\%    & 37.04 & 63.5  & 0.51  & 20000 & 18.5 & 6.78  & 0.32 \\
\hline
\end{tabular}
\caption{\com{Rheological properties and surface tension $\sigma$ of the 75/25\% by weight water/glycerol solvent, and PEO solutions of various mass concentrations ($c = 0.01\%$ to $c = 1\%$). $c^* = 0.291 {\rm kg/m^3}$ or 0.027\% is the critical overlap mass concentration. $\lmr$ is the longest relaxation time of the polymer. $\eta_{\rm 0}$, $\eta_{\rm \infty}$, $\lambda$, and $n_{\rm cf}$ are the parameters obtained from the Carreau fit shown in Fig. \ref{fgr:figure2}(a)}}
\label{tbl:Table1}
\end{table*}

To elucidate the influence of elastic stresses, we use polymer solutions that are weakly shear-thinning, except at large concentrations but with large enough relaxation times for all concentrations studied. The paper is organized as follows: in section \ref{sec:experimental_methods}, we describe the experimental methods and the rheology of the polymer solutions used. We characterize the shear thinning and the relaxation time of the polymer solutions. In section \ref{sec:results}, we compare our results with Newtonian fluids to the results previously reported in the literature. We then present the thinning obtained for the polymer solutions. We observe that different concentrations of polymers lead to different pinch-off dynamics. We discuss our results in section \ref{sec:discussion}, where we describe the influence of elastic forces on the shape of the bubble near the pinch-off. We further highlight this by presenting the contours of the bubble in the polymer solution as it breaks off.


\section{Experimental Methods}
\label{sec:experimental_methods}

The bubble pinch-off experiments are performed in a $50 \times 50 \times 50 \, \milli\meter^3$ transparent tank that contains the liquid (Fig. \ref{fgr:figure2}). A syringe pump (KDS Legato 110) extrudes air bubbles through a stainless-steel nozzle (inner diameter $\ho = 2.31 \, \milli\meter$) at a controlled flow rate $Q$. The vertical alignment of the needle is obtained by using a custom 3D printed setup to place the needle in the tank. The tip of the needle is at least $10\,\milli\meter$ below the air/liquid interface, ensuring that the free surface has no effect on the pinch-off dynamics. We use flow rates in the range $Q = 0.02\,\milli\liter\,\minute^{-1}$ to $0.2 \,\milli\liter\,\minute^{-1}$ depending on the viscosity of the liquid. The flow rate is well below the critical flow rate where the flow transitions to jetting and allows to generate a single bubble of constant volume in a quasi-static regime.\cite{oguz_dynamics_1993} We ensure that changing the flow rate in this range does not influence the results, which confirms that we remain in the quasi-static regime (see supplementary materials).  

\begin{figure*}[ht]
  \begin{center}
    \includegraphics[width=0.97\linewidth]{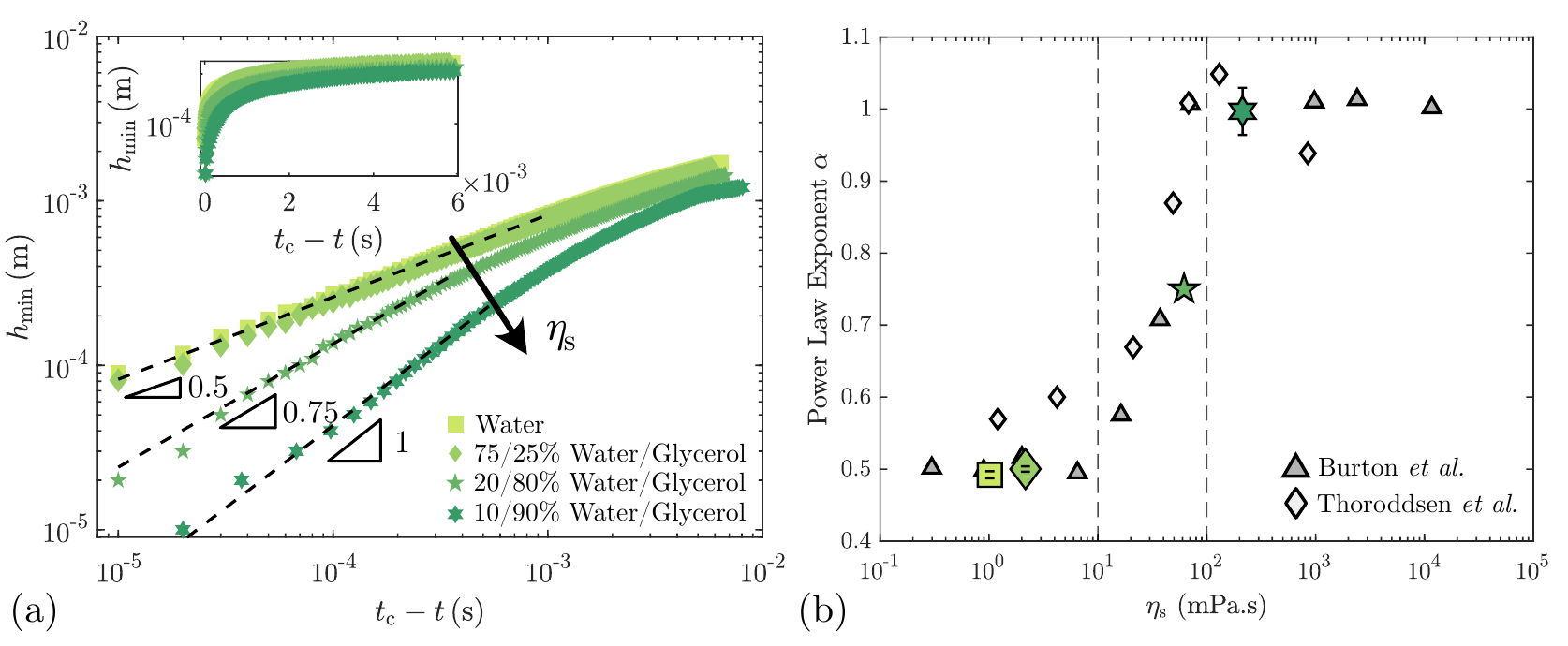}
  \end{center}
  \caption{Bubble pinch-off in Newtonian liquids. (a) Thinning dynamics for different Newtonian liquids: water (${\eta_{\rm s}}=1\,\milli\pascal\,\second$) and mixture of water/glycerol at 75/25\% by weight ($\etas=2.14\,\milli\pascal\,\second$), 20/80\% by weight ($\etas=62.1\,\milli\pascal\,\second$), and 10/90\% by weight ($\etas=213.3\,\milli\pascal\,\second$). The pinch-off occurs at $t=\tc$, and the time goes from right to left. All experiments exhibit a power-law before the pinch-off given by $\hmin = A(\tc-t)^\alpha$. Inset: Semi-log plot of the temporal evolution of the minimum thickness for the same liquids. (b) Evolution of the power-law exponent for air bubbles generated in Newtonian liquids of different shear viscosity. The viscosity is tuned by using water/glycerol mixtures with different compositions by weight of glycerol (between 0\% and 90\%). The exponents compare well with values previously reported in the literature (grey symbols).\cite{burton_scaling_2005,thoroddsen_experiments_2007} The two vertical lines delimit the regime at low viscosity ($\eta_s \leq 10\,{\rm mPa\,s}$) where $\alpha \simeq 0.5$ and at large viscosity where $\alpha \simeq 1$ ($\eta_s \geq 100\,{\rm mPa\,s}$).}
  \label{fgr:figure4}
\end{figure*}

The liquids consist of mixtures of deionized (DI) water and glycerol (Sigma-Aldrich) for the solvent, and the polymer solutions are prepared using Polyethylene oxide (PEO) of molecular weight $M_{\rm w} = 4000\,\rm kg/mol$ (Sigma Aldrich). We tune the viscosity of the solvent $\etas$ by varying the weight fraction of glycerol from $\etas = 1\,\milli\pascal\,\second$ ($0\%$ glycerol),  to $\etas = 213.3\,\milli\pascal\,\second$ ($90\%$ glycerol per weight). The change in the fraction of glycerol has minimal influence on the surface tension $\sigma$.\cite{takamura2012physical} We prepare the polymer solutions by adding the polymer powder to a 75/25\% by weight water/glycerol mixture, slowly mixing them on a roller mixer for 24-48 hours, \com{and we ensured that the polymer solutions are homogeneous at the end of the preparation}. We use PEO of mass concentrations between $c = 0.01\%$ and $c = 1\%$. \com{The surface tension, measured using the pendant drop method, does not vary significantly in this range of concentrations. The measurements are summarized in Table \ref{tbl:Table1}.} An increase in the polymer concentration results in large variations of the shear viscosity $\eta$ and the relaxation time $\lmr$. The shear viscosity is measured using a 50 $\milli\meter$ $1^{\rm o}$ smooth cone and plate geometry on an MCR 302 rheometer (Anton Paar). We report the evolution of the shear viscosity with shear rate $\gmd$ in Fig. \ref{fgr:figure3}(a). We observe that up to a polymer concentration of approximately $c = 0.1\%$, the viscosity remains more or less constant, whereas the largest concentrations exhibit shear-thinning similar to observations reported in past studies.\cite{dekker_supplementary_nodate} The shear viscosity $\eta(\gmd)$ can be fitted using a Carreau model defined as $\eta = \eta_{\rm \infty} + (\eta_{\rm 0} - \eta_{\rm \infty})(1+(\lambda\gmd)^2)^{\frac{n-1}{2}}$.\cite{Bird1987} The fitting parameters $\eta_{\rm 0}$, $\eta_{\rm \infty}$, $\lambda$, and $n$ are the zero-shear viscosity, infinite shear viscosity, a time constant, and the power-law index, respectively. \com{Their values are summarized in Table \ref{tbl:Table1}.} \finalCom{The Carreau model was used to fit the viscosities instead of the Carreau-Yasuda model since it provided a more robust fit for all concentrations of polymer solutions used in the present study. While the Carreau-Yasuda model also fits the data well, it yields a larger error for $c = 1\%$ concentration due to a bias of the fitting model for the large viscosities measured at lower shear rates.} In Fig. \ref{fgr:figure3}(b), we show the longest relaxation times $\lmr$ of the polymer solutions, obtained from droplet pinch-off experiments.\cite{Deblais2020,Rajesh2022} We measure this relaxation time by fitting the rescaled minimum thickness $\hmin/\ho$ during the formation of a liquid droplet in the viscoelastic regime ($t-t_{\rm c} > 0$) with an exponential thinning model \com{$\hmin \propto e^{-(t-\tc)/3\lmr}$}, as shown in the inset of Fig. \ref{fgr:figure3}(b).\cite{Anna2001} The relaxation time varies from $\lmr = 0.02\,\second$ to $\lmr = 0.5\,\second$ when increasing the PEO concentration from $c=0.01\%$ to $c=1\%$. \finalCom{The relaxation time was also measured from the relaxation modulus of the polymer solution for $c = 1\%$ concentration. The measured relaxation time was found to be comparable to the $\lmr$ measured from droplet pinch-off.\cite{varma2022rheocoalescence} (see supplementary material)} The evolution of the relaxation time with $c$ is well captured by the empirical law $\lmr \propto (c/c^*)^{0.66}$. \cite{Tirtaatmadja2006, Deblais2020} Here $c^*$ is the critical overlap concentration above which the polymer coils starts to overlap. For the solutions used here, we calculate $c^* = 0.027\%$.\cite{Graessley1980}

We record the bubble pinch-off using a high-speed camera (Phantom VEO 710) equipped with a macro lens (Nikon Micro-NIKKOR 200 mm f/4 AI-s) and a microscope lens (Mitutoyo X2). The spatial resolution is about $10\,\micro\meter\,\, {\rm per \, pixel}$. The recordings are typically made at 100,000 frames per second or higher, and the setup is back-lit with an LED panel (GSVitec) with a diffuser. Most of the results reported here are obtained at 100,000 frames per second. We ensured that the power-law exponents are independent of the frame rates within the experimental resolution (see supplementary materials). We then process the recordings using custom ImageJ macros and python routines to extract the temporal evolution of the minimum thickness $\hmin$ and outline of the bubble.


\begin{figure*}[ht]
  \begin{center}
    \includegraphics[width=1\linewidth]{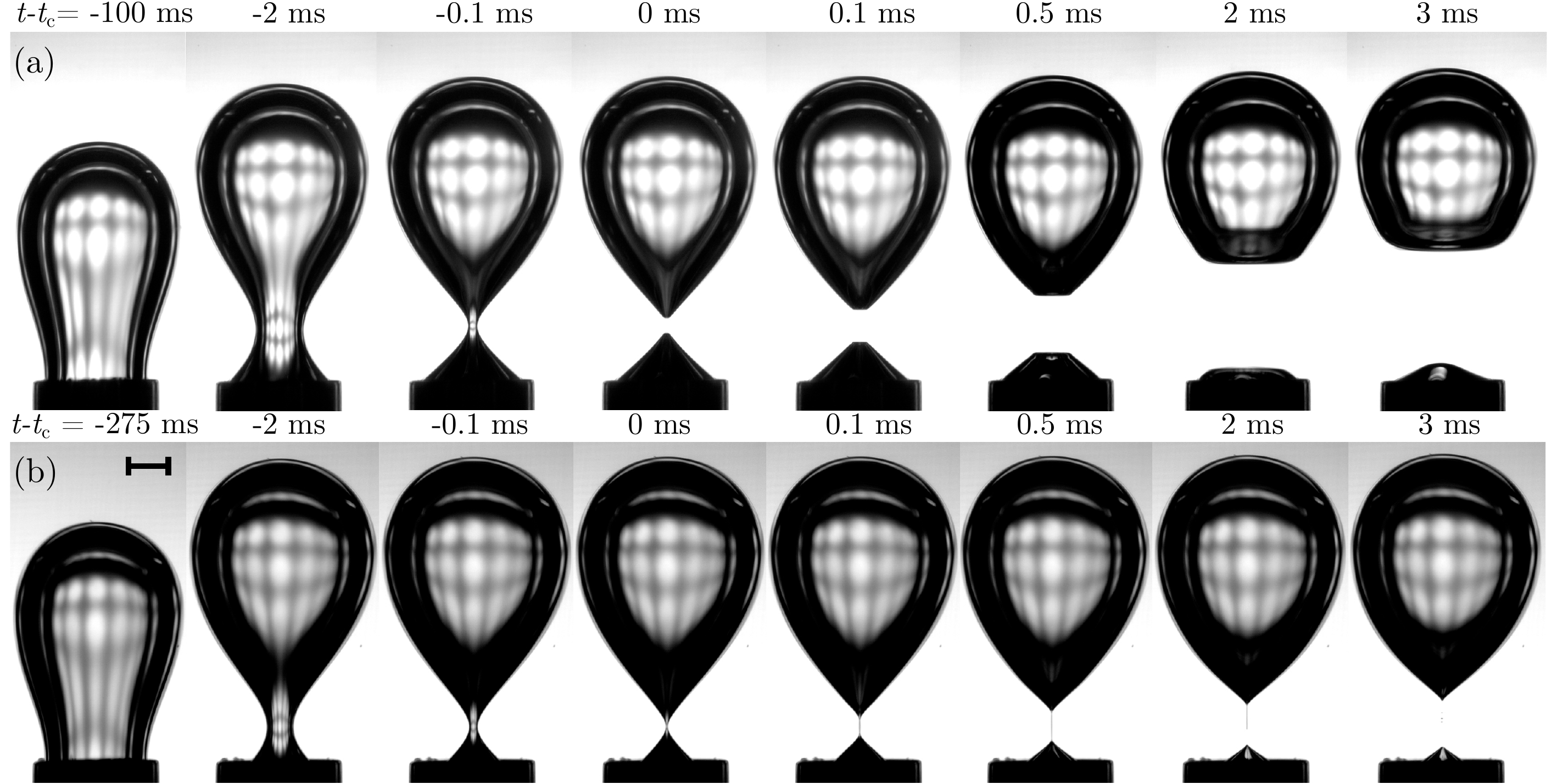}
  \end{center}
  \caption{Time sequence of bubble pinch-off in polymer solutions (a) in a $c=0.01\%$ mass concentration of 4000K PEO in a 75/25\% weight water/glycerol solvent and (b) in a $c=1\%$ mass concentration of 4000K PEO in a 75/25\% weight water/glycerol solvent. The scale bar is 1 mm.}
  \label{fgr:figure5}
\end{figure*}


\section{Results} 
\label{sec:results}

\subsection{Bubble pinch-off in Newtonian fluids}

\com{Because of the short time scales involved in the pinch-off, finding the exact value of $\tc$ is experimentally challenging. We can detect the last frame where the bubble is still connected to the nozzle and the first frame after the pinch-off, but cannot describe the thinning in between due to the limit of our temporal resolution ($10\,\mu{\rm s}$). We numerically improve this accuracy by searching for a best-fit power law in a narrow time range near $\tc$ (see supplementary materials for more details).\cite{dekker_2022} We test this method by comparing our measurements to the values in the literature for Newtonian solvents of different viscosities.\cite{burton_scaling_2005,thoroddsen_experiments_2007}}

\com{In Fig. \ref{fgr:figure4}(a), we report the time evolution of the minimum thickness $\hmin$ for the bubble in solvents with increasing fractions of glycerol, and hence, increasing viscosity. Because of the short timescale of the pinch-off, it is necessary to appropriately define $\tc$ to recover the correct power-law exponent. Our experiments show that for all the viscosities, the resulting thinning dynamics are well-fitted with the equation $\hmin = A(\tc-t)^{\alpha}$. For pure water and the 75/25\% by weight water/glycerol solvent, which are in the inviscid limit ($\etas < 10\,\milli\pascal\second$), the fitting parameters $\alpha \simeq 0.5$ and $A \simeq 0.025$ matches the results of Burton \textit{et al.}\cite{burton_scaling_2005}} \com{In the viscous limit ($\etas > 100\,\milli\pascal\,\second$), the minimum thickness follows $\hmin = (\sigma/\etas)(\tc-t)$.\cite{Eggers2017} The experimental fit leads to $\alpha \simeq 1$,  similar to exponents observed in previous studies.\cite{burton_scaling_2005,thoroddsen_experiments_2007} For the prefactor, we recover $A\simeq 0.4$, which is slightly larger than $\sigma/\etas = 0.34$ for the viscous solvent we use. We note a similar discrepancy in the data reported by Burton \textit{et al.}\cite{burton_scaling_2005} For the 20/80\% by weight water/glycerol solvent , which has an intermediate viscosity $10 \,\milli\pascal\,\second< \etas < 100 \,\milli\pascal\,\second$, the fitting parameters are $\alpha \simeq 0.75$, and $A \simeq 0.135$. In Fig. \ref{fgr:figure4}(b), we summarize $\alpha$ and compare them with the exponents reported by Burton \textit{et al.}\cite{burton_scaling_2005} and Thoroddsen \textit{et al.}\cite{thoroddsen_experiments_2007}} The exponents we recover from our experimental method for the Newtonian solvents (mixture of water and glycerol) match the results observed in the literature. In the following, we consider the influence of viscoelasticity by adding polymers (PEO) to the solvent, and in particular, on the evolution of the thinning dynamics with the polymer concentration.

\subsection{Bubble pinch-off in polymer solution}


\begin{figure*}[ht]
  \begin{center}
    \includegraphics[width=1\linewidth]{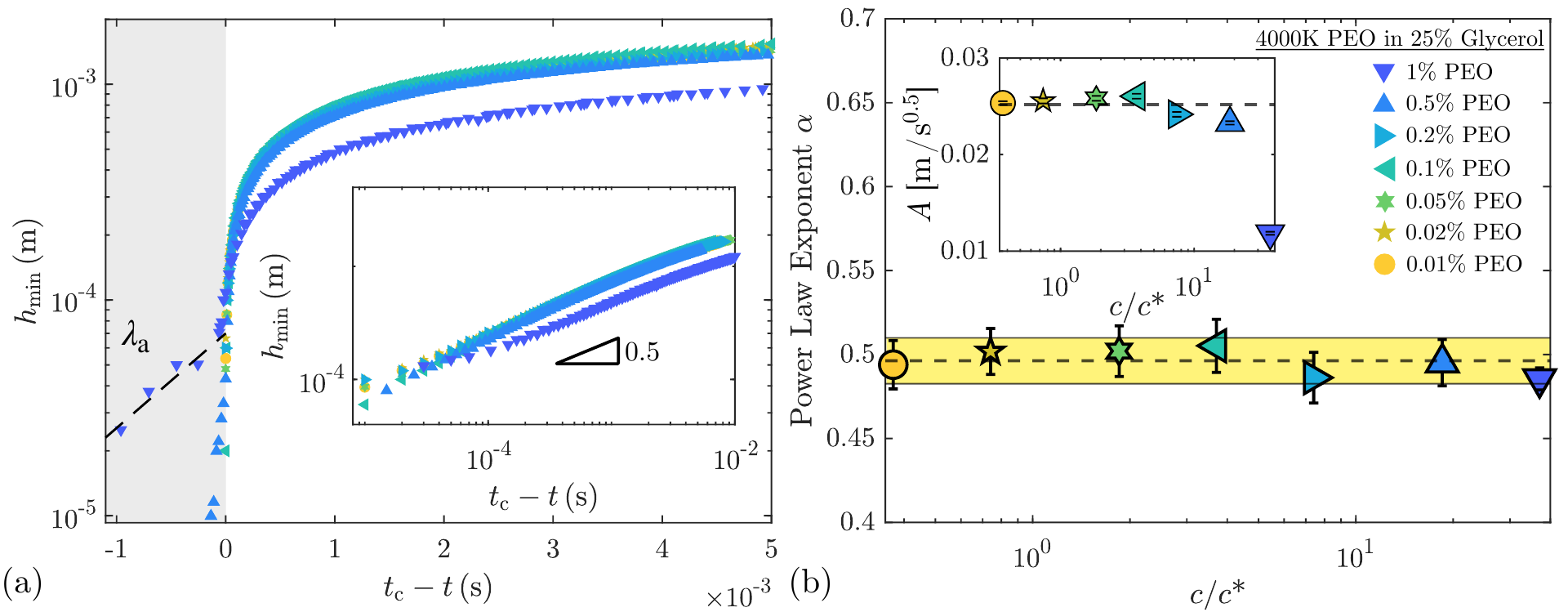}
  \end{center}
  \caption{Evolution of the minimum thickness $\hmin$ of an air bubble in a 4000K PEO solution at different concentrations $c$ in a 75/25\% by weight water/glycerol solvent. Note here that for solutions with a well-defined viscoelastic regime, $\tc$ corresponds to the moment where the pinch-off would have occurred if there were no polymer in the solution. For $\tc-t < 0$, the thinning is fitted with an exponential thinning model with a fitting parameter $\lma$. Inset: Log-log evolution of $\hmin$ for solutions of different concentrations. (b) Exponents $\alpha$ extracted from the power-law $h_{\rm min} = A(t_{\rm c}-t)^\alpha$ for the experiments performed in the polymers solution of rescaled concentrations $c/c^* = 0.37$ to $c/c^* = 37$, corresponding to $c = 0.01\%$ to $c = 1\%$. The horizontal dashed line indicates the exponent $\alpha=0.496 \pm 0.01$. The yellow shaded region is the standard deviation of the measured exponents. Inset: Evolution of the prefactor $A$ with the polymer concentration.}
  \label{fgr:figure6}
\end{figure*}

\begin{figure}[ht]
  \begin{center}
    \includegraphics[width=1\linewidth]{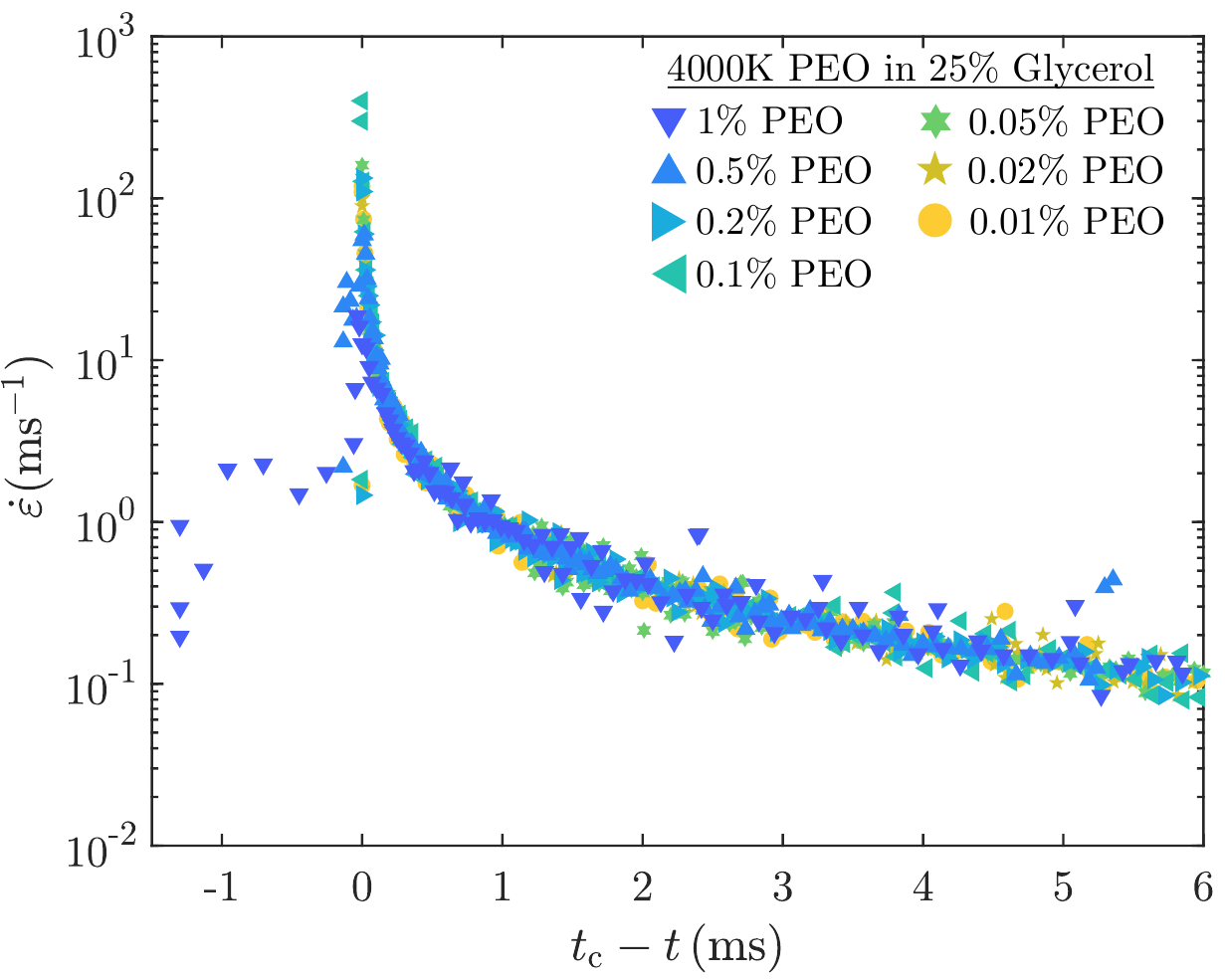}
  \end{center}
  \caption{Evolution of the strain-rate $\eps$ of the thinning for polymer solutions at different concentrations $c$. The strain rate reaches a maximum $\eps_{\rm max}$ when the flow transitions from the Newtonian to the viscoelastic regime.}
  \label{fgr:figure7}
\end{figure}


We now consider the formation of an air bubble in a viscoelastic solution of polymers. As shown in Figs. \ref{fgr:figure1}(c) and \ref{fgr:figure1}(d), the pinch-off of a drop of Newtonian liquid and a polymer solution in air are drastically different. Indeed, the thinning and pinch-off of a polymer solution droplet in air exhibits two successive regimes.\cite{Rajesh2022} First, a Newtonian regime where the thinning of the solution is similar to the solvent in which it was prepared, captured by $\hmin \propto [\sigma(\tc-t)^2/\rho]^{1/3}$. Here $\tc$ is the time at which the drop would have broken if there were no polymers dissolved.\cite{ThievenazSuspensions,Rajesh2022} The presence of polymers results in an increase in the extensional viscosity as the liquid thins, and delays the pinch-off.\cite{Amarouchene2001} Around $t=\tc$, the thinning slows down due to the coil-stretch transition of the polymers, and becomes viscoelastic.\cite{ThievenazSuspensions,Rajesh2022} The viscoelastic regime is characterized by a thread connecting the drop to the nozzle. This ligament thins exponentially as $\hmin \propto e^{-t/(3\,\lmr)}$ where $\lmr$ is the longest relaxation time of the polymer in the solvent.\cite{Anna2001}


For droplet pinch-off, the presence of polymer is felt even at very low concentrations, such as 0.001\% or 10 parts per million by weight for solutions of a 4000K PEO.\cite{Deblais2020} Bubble pinch-off in polymer solution is different. In Figs. \ref{fgr:figure5}(a)-(b), we illustrate the thinning of an air bubble in polymer solutions of two different concentrations. Fig. \ref{fgr:figure5}(a) illustrates the formation of an air bubble in a solution of 4000K PEO at $c=0.01\%$ and Fig. \ref{fgr:figure5}(b) at $c=1\%$ mass concentration prepared in a 75/25\% by weight water/glycerol solvent. For a certain duration of time, the thinning of the bubble appears similar in solutions of both low and high concentrations. At low polymer concentrations ($c \leq 0.02\%$), the bubble pinch-off is similar to the inviscid solvent. Following the definition for the solvents, the bubble detaches at time $t=\tc$. Above a certain concentration ($c \geq 0.05\%$), the pinch-off is modified; an air thread appears that binds the bubble to the nozzle (see supplementary materials). \com{For bubbles in polymer solutions of concentration $c \geq 0.05\%$, we define $t=\tc$ as the moment when the thin and slender air thread first appears.} This structure lasts for a very short time before the pinch-off. The thickness and lifetime of this structure increase with concentration, similar to the viscoelastic thinning of a drop in air. At the highest concentrations considered here ($c = 0.5\%$ and $c = 1\%$), we have sufficient spatial and temporal resolutions to quantify this thinning, as shown in Fig. \ref{fgr:figure6}(a).


\begin{figure*}[ht]
  \begin{center}
    \includegraphics[width=0.95\linewidth]{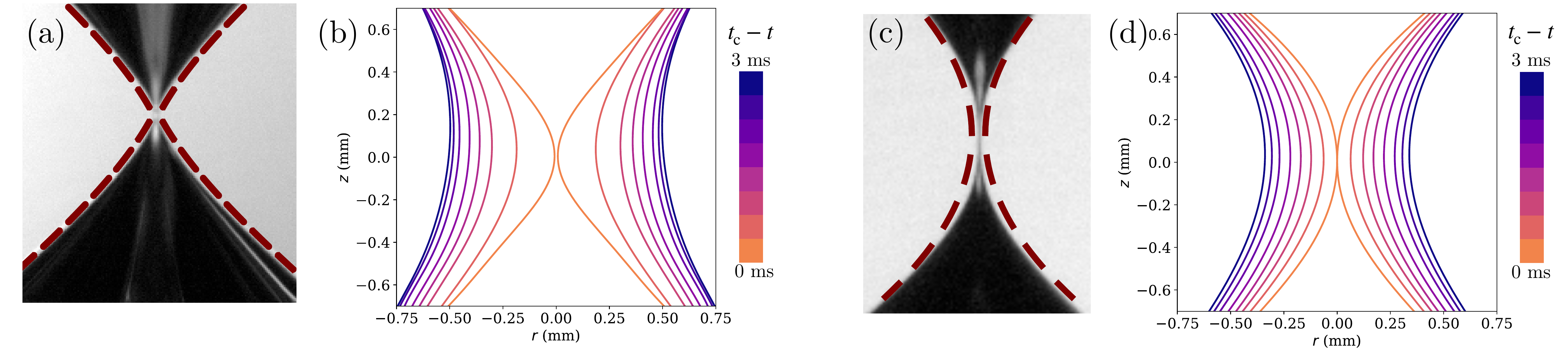}
  \end{center}
  \caption{Bubble neck profile and spatial evolution for Newtonian liquids. (a) A close-up view of the neck near the pinch-off in a 75/25\% by weight water/glycerol (inviscid) solvent. The dotted lines at the liquid/air interface is a hyperbolic fit. (b) Polynomial fitted spatial evolution of the bubble in the corresponding solvent. (c) The bubble near the pinch-off in a 10/90\% water/glycerol (viscous) solvent. The dotted line is a parabolic fit. (d) Polynomial fitted spatial evolution of the neck in the viscous solvent.}
  \label{fgr:figure8}
\end{figure*}


\begin{figure*}[ht]
  \begin{center}
    \includegraphics[width=0.95\linewidth]{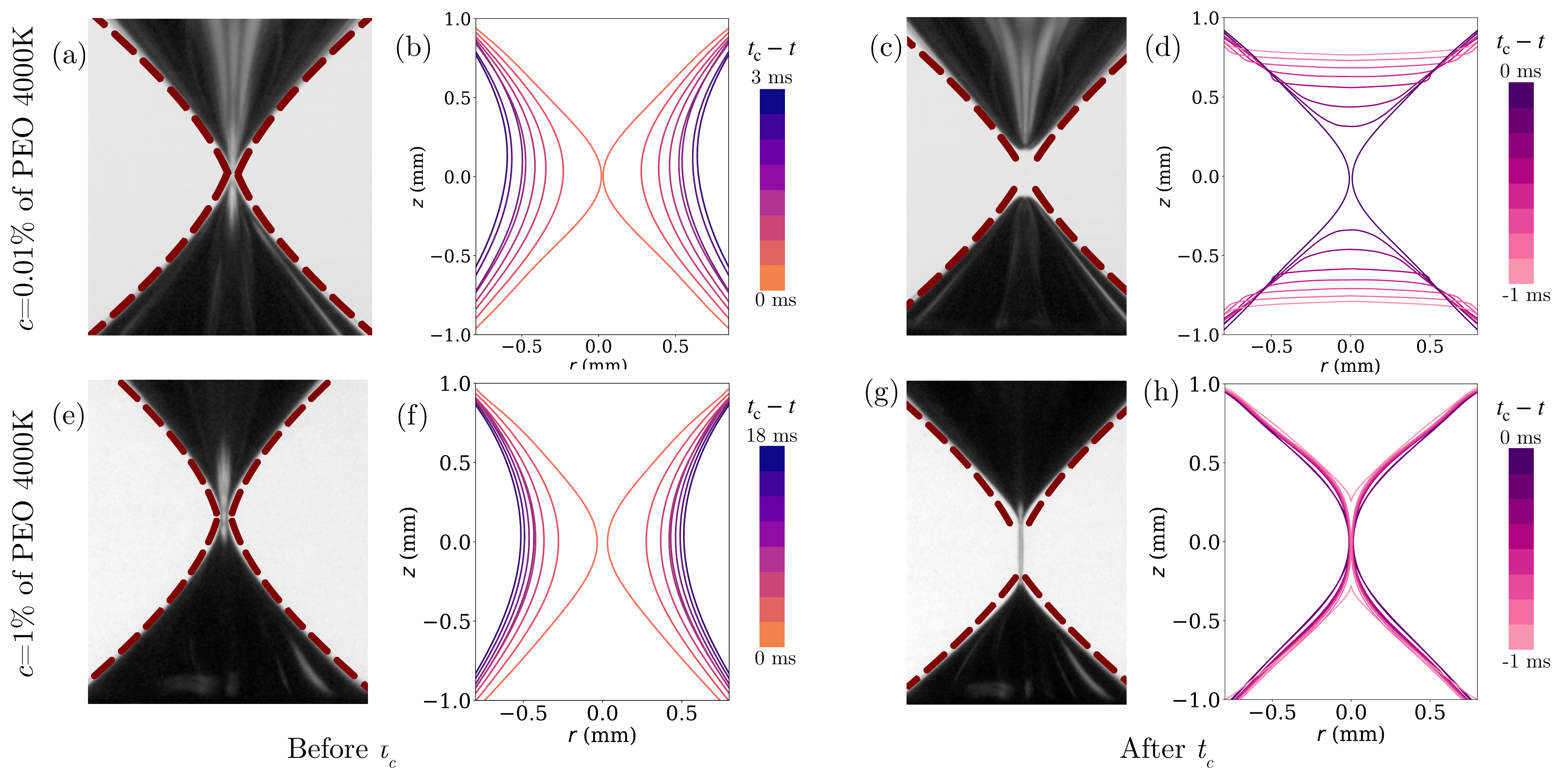}
  \end{center}
  \caption{Bubble neck profile and spatial evolution of the neck in 4000K PEO solution in a 75/25\% by weight water/glycerol solvent and (a)-(d) $c=0.01\%$, (e)-(h) $c=1\%$. (a) The bubble neck near the pinch-off, with a hyperbola fit. (b) Polynomial fitted spatial evolution of the neck near the pinch-off (c) The neck a $\micro\second$ after the pinch-off, with a hyperbolic fit. (d) Polynomial fitted spatial evolution of the neck after the pinch-off. (e) The bubble neck a few $\micro\second$ before the air thread forms at $\tc -t = 0$, fitted with a hyperbola. (f) Polynomial fitted spatial evolution of the neck before the air thread forms. (g) The bubble neck a $\micro\second$ after the pinch-off. The ends of the air thread is fitted with a hyperbola. (h) Polynomial fitted spatial evolution of the air thread.}
  \label{fgr:figure9}
\end{figure*}


Similarly to the Newtonian liquid, we extract the time evolution of the minimum neck thickness $\hmin$ for polymer solutions of 4000K PEO of mass concentration 0.01\% to 1\% in Fig. \ref{fgr:figure6}(a). As noted before, and similar to the case of the pinch-off of polymer solutions droplet in air,\cite{Rajesh2022} when the air thread appears, we define $\tc$ as the time of its appearance, rather than the pinch-off time. While the difference is not strongly noticeable for dilute polymer solutions, the distinction must be made for larger concentrations (for solutions of $c \geq 0.05\%$). Indeed, for instance, for $c \geq 0.5\%$ of PEO, the pinch-off of the bubble only happens a few ms after the thread-like structure first appears. This is visible as a new regime of thinning in Fig. \ref{fgr:figure6}(a) when $\tc-t < 0$. Using the above criteria to select $\tc$, we report the time evolution of $\hmin$ in log-log scale in the inset of Fig. \ref{fgr:figure6}(a). Interestingly, all the polymer solutions considered in this study follows the same power law, $\hmin = A(t-\tc)^\alpha$ with $\alpha = 0.496 \pm 0.01$. This result is remarkable, especially for solutions of higher concentrations, since the zero-shear viscosity of solutions spans over four orders of magnitude. \com{We further note that the prefactor $A \simeq 0.025$ remains equal to that of the solvent, up to a concentration c = 0.1\%, and then decreases. This captures the slowing down of the thinning of the higher concentration solutions.} We also draw attention to the thinning of the air thread when $\tc - t < 0$. At the highest concentrations we study ($c = 0.5\%$ and $c = 1\%$ mass concentration in 4000K PEO in 75/25\% by weight water/glycerol solvent), we characterize the thinning using an exponential thinning model.\cite{Anna2001} When $\tc-t<0$, we define the minimum thickness as $\hmin \propto e^{{-(\tc-t)}/{(3\,\lma)}}$. Here, $\lma$ is a fitting parameter for the thinning of the air thread surrounded by a polymer solution, and is different from the relaxation time $\lmr$ of the polymer solution. We estimate the values of this parameter as $\lma \sim 0.027\,\milli\second$ and $\lma \sim 0.33\,{\rm m}\second$ for $c = 0.5\%$ and $c = 1\%$ concentrations, respectively. We note that the values are at least three orders of magnitude smaller than the relaxation times $\lmr$. This suggests that the pinch-off of a polymer drop in air and a gas bubble in a polymer solution, although qualitatively similar, have notable differences.



\section{Discussion}
\label{sec:discussion}


The thinning of a bubble in a polymer solution has different behaviors depending on the solution concentration. However, by appropriately selecting $\tc$, we can extract the value of the power-law exponent $\alpha$ in the Newtonian regime. The log-log inset in Fig. \ref{fgr:figure6}(a) shows the evolution in the power-law regime for $c = 0.01\%$ to $c = 1\%$ mass concentration. We summarize the value of the exponent $\alpha$ when varying $c$ in Fig. \ref{fgr:figure6}(b) and obtain \com{$\alpha = 0.496 \pm 0.01$} for all polymer concentrations considered here. The results we report here, where the power-law exponent is independent of the polymer concentration, are similar to a recent study on the singular coalescence of polymer solutions.\cite{dekker_2022} For solutions that show weak shear thinning and strong viscoelasticity, similar to the solutions we use, the elastic coalescence also modifies the interface shape. As discussed later, we see a similar strong modification of the bubble interface near the pinch-off. Furthermore, the result is similar to the droplet pinch-off of polymer solutions, where the Newtonian regime of the thinning has an exponent $\alpha = 2/3$ independent of the concentration.\cite{ThievenazSuspensions,Rajesh2022} We discuss two possible flow behavior of the external liquid during the bubble pinch-off to qualitatively rationalize why the exponent remains independent of the concentration. We assume that during the thinning, the liquid viscosity is either purely shear-thinning, or extensional thickening. \com{In the inset of Fig. \ref{fgr:figure6}(b), we observe that the prefactor $A$, obtained from the best fit of a power-law, decreases with the polymer concentration, and has a sharp drop for $c = 1\%$ concentration. Similar observations have been made during the droplet thinning of solvents and suspensions of increasing viscosity.\cite{ThievenazSuspensions} Therefore, the prefactor can hence be considered as an amplitude of the second-order effect of viscosity on the thinning. This decrease in $A$, and the observed slowing down of the thinning with increase in concentration suggests that the extensional thickening is a more likely mechanism at play here.}

For the thinning shown in Fig. \ref{fgr:figure6}(a), we show the strain-rate $\eps$ near the liquid/air interface in Fig. \ref{fgr:figure7}. The strain-rate scales as $\eps \sim \dot{h}_{\rm min}/\hmin$.\cite{Amarouchene2001} For the solvent and low concentration solutions, $\eps$ diverges near $\tc$. At larger concentrations \com{corresponding to the formation of the air thread}, $\eps$ reaches a maximal value near $\tc$ and decreases when $\tc -t < 0$. This decrease is clearly visible for polymers concentrations of $c = 0.5\%$ and $c = 1\%$, where the experimental resolution allows us to track the thin air thread. In a shear flow, the polymer solutions are shear thinning, as shown in Fig. \ref{fgr:figure3}(a). If the solution shear thins, the thinning near the pinch-off would become faster. However, the thinning slows down near the pinch-off in polymer solutions. A slower thinning suggests an increase in the flow resistance, similar to an extensional viscosity thickening in droplet pinch-off.\cite{Amarouchene2001} The increase in the flow resistance in the droplet pinch-off arises due to a coil-stretch transition when the flow reaches a critical strain rate.\cite{DeGennes1974} Near $t \rightarrow \tc$, the strain-rate increases sharply, as shown in Fig. \ref{fgr:figure7}. Hence, it is likely that $\eps(\tc-t\rightarrow0)$ is large enough to uncoil the polymers in the vicinity of the liquid-air interface due to the local elongational flow. A macroscopic manifestation of the elastic interaction of the uncoiled polymers with the solvent is the modified interface shape we observe at larger concentrations, as can be seen when $t-\tc > 0$ in Fig. \ref{fgr:figure5}(b). When $\tc-t > 0$, far from the pinch-off, $\eps$ is below the coil-stretch transition limit and the polymers do not interact with the flow.\cite{Rajesh2022,ThievenazSuspensions} The thinning depends only on the solvent, as suggested by Fig. \ref{fgr:figure6}(a) where the evolution of $\hmin$ for different polymer concentrations overlaps. This is further confirmed by the recovered exponents, which are independent of the polymer concentration. The slight difference observed at the highest concentration is associated with the increased viscosity due to the spherical polymer coils, which follows Einstein’s laws.\cite{Mardles1940} 

When the strain-rate $\eps$ of the flow reaches its maximal value near $t = \tc$, the polymers likely start to uncoil locally near the interface. Further, the maximum coil-stretch transition strain-rate $\eps_{\rm max}$ decreases with the polymer concentration.\cite{Rajesh2022} There are two possible explanations for why we see the air thread only above a threshold concentration. Indeed, it is likely that $\eps_{\rm max}$ for the low concentration solutions ($c = 0.01\%$ and $c = 0.02\%$) is larger than the time resolution of our experiments, and the air thread thickness smaller than the lowest pixel length. Alternatively, the time scale of the singularity may be larger than the time scale of the coil-stretch transition strain-rate ($1/\eps_{\rm max}$) of the solution, hence the bubble pinches off before it enters the viscoelastic regime.


The droplet and bubble pinch-off have similar features, such as a period of Newtonian power-law thinning, which is independent of the concentration, followed by a transition to a viscoelastic regime. However, we highlight certain key differences in the thinning, particularly during the viscoelastic regime. Although the air thread in the bubble pinch-off in polymer solutions resembles the viscoelastic filament in the droplet pinch-off, the thread itself has no polymers. The thinning of this thread is likely governed by the stretching of the polymers in the vicinity of the liquid/air interface. We recall that the air threads in the solutions with mass concentrations $c = 0.5\%$ and $c = 1\%$ have sufficient thickness to characterize their thinning. Although $\lma$, the fitting parameter we obtain from the exponential thinning,\cite{Anna2001} resembles the relaxation time $\lmr$ of the droplet thinning, they are at least three orders magnitude smaller. \com{Further, the prefactor of the thinning, $A$, decreases with increasing concentration. This suggests a slower thinning of the higher concentration solutions during the pinch-off. We also estimate the time-scale of the stretching of the $c=0.5\%$ and $1\%$ polymer solutions during the bubble pinch-off as $1/\eps_{\rm max} \sim 0.03$ ms. This gives an order of magnitude estimate of the extensional viscosity as $\eta = 2\sigma/(\hmin\,\eps_{\rm max}) \sim 80$ mPa s. It is also interesting to note that by rearranging the expression for the extensional viscosity $\eta = 2\sigma/(\hmin\,\eps_{\rm max})$,\cite{Amarouchene2001} we obtain $\tau_{\rm \eta,c}\sim 1/\eps_{max}$, the visco-capillary time scale.} The observations here is of an apparent paradox; even though the higher concentration solutions are more viscous, the exponent we recover is of an inviscid pinch-off.



Singularities in the bubble pinch-off are regions of topological transitions, where the initial bubble separates to form a main bubble and micron-sized satellite bubbles.\cite{Gordillo_Satellite_Bubbles2007} The differences in the topology arise either as a function of viscosity in Newtonian liquids, illustrated in Figs. \ref{fgr:figure8}(a)-(d), or elasticity, which is the case in polymer solutions shown in Figs. \ref{fgr:figure9}(a)-(h). In Fig. \ref{fgr:figure8}(a), we show the bubble neck in a 75/25\% by weight water/glycerol solvent before the pinch-off. The neck is fitted with a hyperbola of the form \com{$r = R_{\rm 1}\sqrt{1+z^2}$,} where $r$ and $z$ are the neck radius and the width, respectively, and $R_{\rm 1}$ is a fitting parameter. Similar fits have been considered for a bubble neck in an inviscid liquid in past studies.\cite{thoroddsen_experiments_2007} Fig. \ref{fgr:figure8}(b) reports the spatial evolution of the neck for the time $\tc -t = -3$ to $0\,\milli\second$ using a polynomial fit. As the thinning approaches the singularity, we see the neck evolving from a smooth profile to a sharp corner before the pinch-off. When the bubble thins in a more viscous solvent, like the 10/90\% water/glycerol shown in Fig. \ref{fgr:figure8}(c), the neck is modified. Viscosity has a smoothing effect on the neck, which is fitted with a parabola $r = az^2+bz+c$, with $a$, $b$, and $c$ as the fitting parameters. The parabolic neck has also been observed in previous studies.\cite{bolanos-jimenez_effect_2009} Fig. \ref{fgr:figure8}(d) shows the polynomial fitted spatial neck profile a viscous liquid for $\tc -t = -3$ to $0\,\milli\second$. In a viscous liquid, the neck has a smooth profile as it approaches the singularity.

For a bubble in polymer solutions, stretching of the polymers in the vicinity of the liquid/air interface results in an elastic contribution. At low concentrations, this contribution is likely beyond the limits of what we observe. Hence, the bubble neck profile for low polymer concentrations is similar to the bubble neck in the solvent. For a 0.01\% mass concentration, we fit the neck near the pinch-off with a hyperbola as illustrated in Fig. \ref{fgr:figure9}(a). As reported in Fig. \ref{fgr:figure9}(b), the spatial evolution of the neck for the time $\tc -t = -3$ to $0\,\milli\second$ is also comparable to the solvent, which evolves at a similar rate. After the pinch-off, the hyperbolic neck in Fig. \ref{fgr:figure9}(c) destabilizes within a few milliseconds. We show this decay in Fig. \ref{fgr:figure9}(d). At a higher concentration, the local Deborah number $De\sim\lmr\eps_{\rm max}$ is of the order of $10^4$. Thus, we expect the elastic forces to influence the bubble neck near the pinch-off. The solution is also more viscous due to a larger concentration of polymers. From a viscosity-based argument alone, we expect a smoother neck near the pinch-off for higher polymer concentration. However, the neck remains hyperbolic, as seen in a solution of 1\% concentration in Fig. \ref{fgr:figure9}(e). We also note that as $t \rightarrow \tc$, the neck profiles are similar to the solvent and 0.01\% solution. However, because of a larger viscosity of the solution, the neck evolves over a time $\tc -t = -18$ ms to $0\,\milli\second$, as reported in Fig. \ref{fgr:figure9}(f). As discussed earlier, the dissolved polymers influence the thinning only when the strain-rate is sufficiently high. For $\tc-t < 0$, the elastic contribution modifies the neck to form the air thread visible in Fig. \ref{fgr:figure9}(g). Interestingly, we note that the neck, which connects the air thread, remains a hyperbola as the thread thins. Fig. \ref{fgr:figure9}(h) reports the fitted thinning of the neck. For $\tc-t < 0$, the solution elasticity strongly modifies the neck profile of the bubble. Hence, up to the point when the neck profile approaches $t = \tc$, the thinning depends only on the viscosity of the solvent. Elasticity begins to influence the thinning around $t = \tc$, which results in a unique bubble neck, different from the neck profiles reported for a bubble in inviscid and viscous liquids. 


\section{Conclusion}
\label{sec:conclusion}

The pinch-off of a bubble is a classic example of singularities in fluids. As the bubble approaches pinch-off, the minimum thickness of the bubble neck follows a power-law. For a bubble in a Newtonian liquid, the power-law exponent $\alpha$ is a function of liquid density; it takes a value $\alpha = 0.5$ for low viscosity liquids and $\alpha = 1$ for high viscosity liquids. For a narrow range of viscosities in between, the exponent lies in the range $0.5 < \alpha < 1$. In this study, we consider the pinch-off of bubbles in polymer solutions of different concentrations. The polymers are almost Newtonian except at the highest concentrations but produce strong \com{viscoelastic} effects in extensional flows at all concentrations. Further, the solutions have zero-shear viscosities varying over four orders of magnitude,  and higher than that of the solvent. Thinning of the bubbles in low concentration polymer solutions (up to $c = 0.02\%$) is similar to that of the solvent. But for sufficiently high concentrations ($c \geq 0.05\%$), an air thread appears for a short time before the pinch-off. By defining an appropriate $\tc$, solutions of all concentrations follow a power-law thinning with exponent $\alpha \simeq 0.5$. Above a large enough concentration ($c > 0.5\%$ here), the spatio-temporal resolution of our experiments allows us to quantify the minimum thickness of the air thread. The thinning here indicates the existence of two distinct regimes before the pinch-off. Similar to the droplet pinch-off of polymer solutions in air,\cite{Rajesh2022} we define $\tc$ not as the time when the bubble separates, but rather as when the thinning transitions from one regime to another. This approach allows us to show that the thinning of a bubble in polymer solutions also follows a power-law with an exponent $\alpha \simeq 0.5$. This result is of notable interest as the solutions at larger concentrations are more viscous. \com{We attempted to rationalize this result by noting that near $t = \tc$, the thinning slows down. This suggests that the viscosity of the solution increases, similar to the extensional thickening observed during the droplet pinch-off of a polymer solution.\cite{Amarouchene2001, Rajesh2022} The increase in the viscosity is likely due to the unwinding of the polymers near the liquid/air interface at a critical strain-rate.\cite{DeGennes1974} Until the point of coil-stretch transition, the polymer remains coiled and does not interact with the flow. This could explain why the exponent observed for the solutions remains similar to the solvent. The uncoiling and interaction of the polymers with the flow is seen as the unique structure formed by the bubble before it pinches off.} The exponent observed also echoes the recent results obtained for the singular coalescence of polymer solution droplets, where a single exponent characterizes the power-law thinning independent of the concentration.\cite{dekker_2022}

The strong elastic effects due to the high polymer concentrations also lead to a change in the topology of the bubble near the pinch-off. For pinch-off in low viscosity Newtonian liquids, the neck of the bubble has a hyperbolic shape. Unlike the high viscosity solvents, which exhibit a parabolic neck, the polymer solutions retain the hyperbolic shape as they approach $t = \tc$. At sufficiently large concentration, a thread-like structure becomes barely visible for a short time as a result of strong elastic forces experienced by the solution at high strain rates. At even higher concentrations, the effect on the topology of the air-liquid interface is evident. After a period of thinning during which the bubble has a hyperbolic shape and power-law thinning, it transitions to a regime with an elongated thread appearing, analogous to the cylindrical structure in the pinch-off of polymer drops. 



\section*{Conflicts of interest}
There are no conflicts to declare.

\section*{Acknowledgements}
This material is based upon work supported by the National Science Foundation under NSF CAREER Program Award CBET Grant No. 1944844.
The authors thank A. Pahlavan for helpful discussions. 


\balance

\bibliography{bubble_pinchoff_main.bib} 
\bibliographystyle{unsrt} 

\end{document}


\maketitle

\vspace{15mm}


\noindent {\Large {\textsc{Defining the moment of Pinch-off}}}

\vspace{5mm}

Due to the nature of the experiments, defining the exact moment of pinch-off $t_{\rm c}$ is challenging. Indeed, we use image processing routines to extract the minimum thickness of the neck connecting the air bubble and the nozzle and are thus limited by this spatial resolution. Besides, we are able to detect the neck up to the last frame before the bubble separates. These spatial and temporal uncertainties introduce some variation on the exact value of $t_{\rm c}$. We use a python routine to manually select a region close to the last data point (which corresponds to the smallest minimum thickness measured before pinch-off). For polymer solutions with $c \geq 0.05\%$, a filament appears between the bubble and the nozzle for a short time before pinch-off. In this case, the routine is slightly modified and will be discussed in the next section. We adopt an approach similar to the one used recently for the coalescence of liquid droplets \cite{dekker2022elasticity}. For the processing, the time of the smallest minimum thickness measured is defined as a temporary value of $t = t_{\rm c}$. We then vary this value $t_{\rm c}-t$ in a narrow time range $\Delta t$ and perform a best fit routine of this time axis with the minimum thickness. The exponent corresponding to the lowest value of the residual obtained from this best fit is the exponent associated with the thinning, which is reported in this study. 


\newpage

\noindent {\Large {\textsc{Modified definition for $t_{\rm c}$ in polymer solutions}}}

\medskip

\begin{figure}[h!]
 \begin{center}
 \includegraphics[width=1\textwidth]{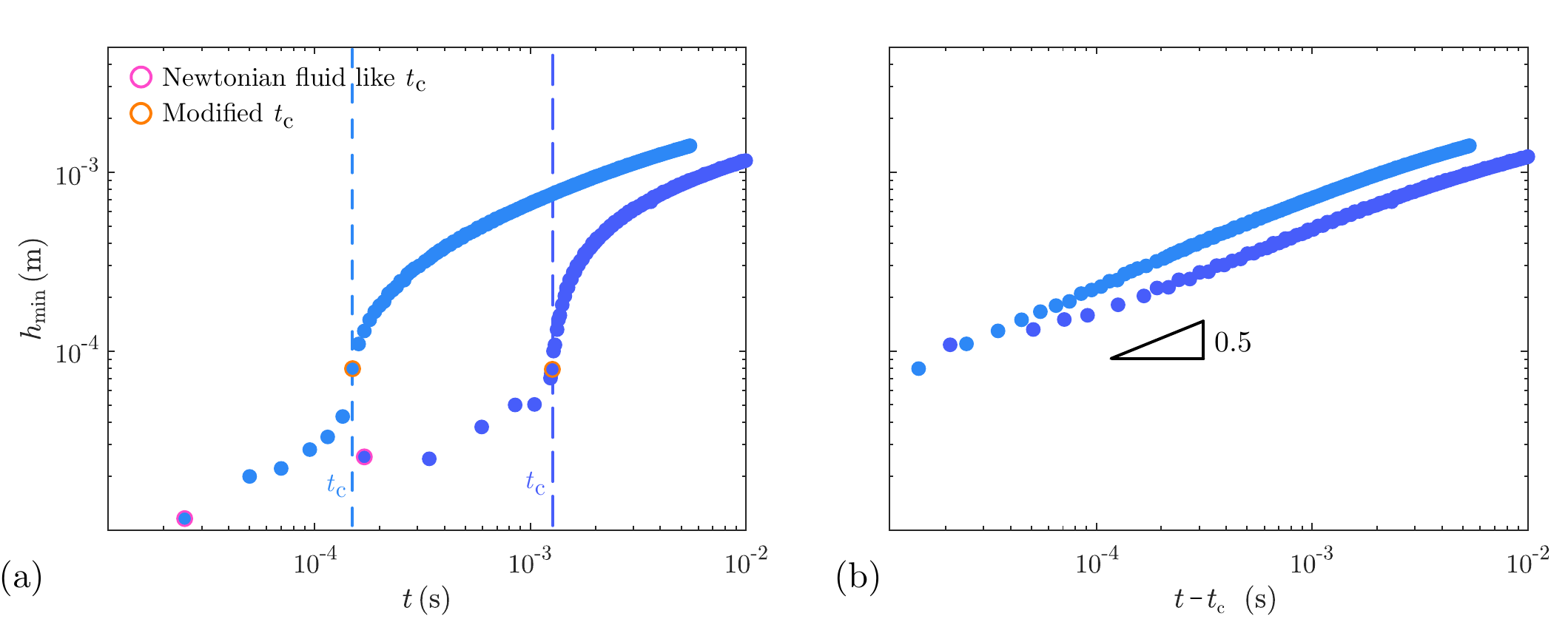}%
 \caption{Thinning of a bubble in polymer solutions of 4000K PEO with mass concentration $c = 0.5\%$ (light blue) and $c = 1\%$ (dark blue) in a 75/25 wt\% water/glycerol solvent at a flow rate $Q = 0.02 \, {\rm mL\, min^{-1}}$. (a) Evolution of the thinning dynamics without applying any time shift. (b) Log-log scale evolution after selecting $t_{\rm c}$ as the instant where the neck transition to an elongated thread.}
 \label{fgr:SM_Fig3}
  \end{center}
 \end{figure}

For solutions of 4000K PEO at concentration $c \geq 0.05\%$, our experiments show that the filament shape is modified before pinch-off. The initial thinning exhibits a Newtonian-like dynamics. At later time, an air thread forms, connecting the bubble to the nozzle. The thinning of this air thread follows a different dynamics. This dynamics is visible in Fig. \ref{fgr:SM_Fig3}(a), which is obtained by using the definition of $t_{\rm c}$ used for Newtonian fluids (shown in pink). We modify the definition of $t_{\rm c}$ by selecting the moment the thread-like structure first appears as the new value of $t_{\rm c}$. A similar approach has been successfully used for the pinch-off of droplets of polymers (see \textit{e.g.}, \cite{rajesh2022transition}). 

This modified evolution of the thinning is shown in Fig. \ref{fgr:SM_Fig3}(b). Using this rescaling of time leads to a power-law evolution of the thinning dynamics in the Newtonian regime. The best fit of the data with the equation $(t_{\rm c}-t)^\alpha$ shows here that we obtain an exponent close to 0.5. Note that such a power law is not recovered if we consider $t_{\rm c}$ as the final time of pinch-off. The power-law evolution is only valid for $t<t_{\rm c}$, \textit{i.e.}, in the Newtonian-like regime.

\medskip

 \newpage

\noindent {\Large {\textsc{Influence of the flow rate and the frame rate on the measured thinning dynamics}}}

\vspace{3mm}

To ensure that the results reported in this study are not dependent on the experimental conditions, we vary the following experimental parameters: 1) the flow rate $Q$ of the syringe pump used to generate the air bubble and 2) the frame rate of the recordings. 

\medskip
We set the syringe pump to the following flow rates $Q = 0.02, 0.05, 0.1, 0.2$ and $0.5\,\rm{mL\,min^{-1}}$. The resulting thinning dynamics for the different flow rates are reported in Fig. \ref{fgr:SM_Fig1}(a). After selecting appropriate values of $t_{\rm c}$, the thinning at different flow rates overlaps well to reveal an exponent $\alpha  \simeq 0.5$. Further, we extract the exponents with best fit power-laws and report the values in the inset of Fig. \ref{fgr:SM_Fig1}(a). For the range of flow rate $Q$ considered, the variations in the exponent is within the accuracy of the experiment, here $\alpha = 0.5094 \pm 0.0129$. These experiments confirmed that the results reported in this study do not depend on the flow rate in the range $Q \leq 0.5 \, {\rm mL \, min^{-1}}$ considered here.

\medskip

We also show in Fig. \ref{fgr:SM_Fig1}(b) the thinning dynamics extracted from movies recorded at frame rates of 75,000, 100,000, 150,000, and 200,000 frames per second. After shifting the thinning with appropriate $t_{\rm c}$ using the method described in the previous section, we observe that the thinning of the neck follows power-law evolution with best fit exponents $\alpha = 0.5023 \pm 0.023$, with no particular trend observed [see inset of Fig. \ref{fgr:SM_Fig1}(b)]. The value of the exponent of the power law does not depend on the frame rate of the recordings for the experimental conditions investigated in the main article, providing that the frame rate used is large enough.

 \begin{figure}[h!]
 \begin{center}
 \includegraphics[width=\textwidth]{figures_sm/Dependance on Flow rate and FPS.pdf}
 \caption{Thinning dynamics of air bubbles in polymer solutions of 4000K PEO with $c = 1\%$ mass concentration in a 75/25 wt\% water/glycerol solvent. (a) Thinning dynamics for different flow rates $Q$ between $0.02\,{\rm mL/min}$ and $0.5\,{\rm mL/min}$. (b) Thinning dynamics recorded at different frame rates between 75,000 and 200,000 frames per second. The insets in both figures show the exponents obtained by the best-fit power law $(t_{\rm c}-t)^\alpha$.}
 \label{fgr:SM_Fig1}
  \end{center}
 \end{figure}

\vspace{10mm}

\newpage


\noindent {\Large {\textsc{Thinning in Newtonian solvents}}}

\medskip

The thinning dynamics in Newtonian fluids depends on the solvent viscosity, leading to power-law evolutions differing for inviscid and viscous solvents. This is shown in the main article, where the exponent $\alpha \simeq 0.5$ for inviscid solvents and $\alpha \simeq 1$ for viscous solvents is recovered \cite{burton2005scaling}. The difference in thinning is also qualitatively visible at the bubble neck, which has a more conical or hyperbolic shape in the inviscid solvent. However, when the viscous forces are dominant, the thinning is slowed down, and the resulting neck can be fitted with a parabola. 

 \begin{figure}[h!]
 \begin{center}
 \includegraphics[width=0.95\textwidth]{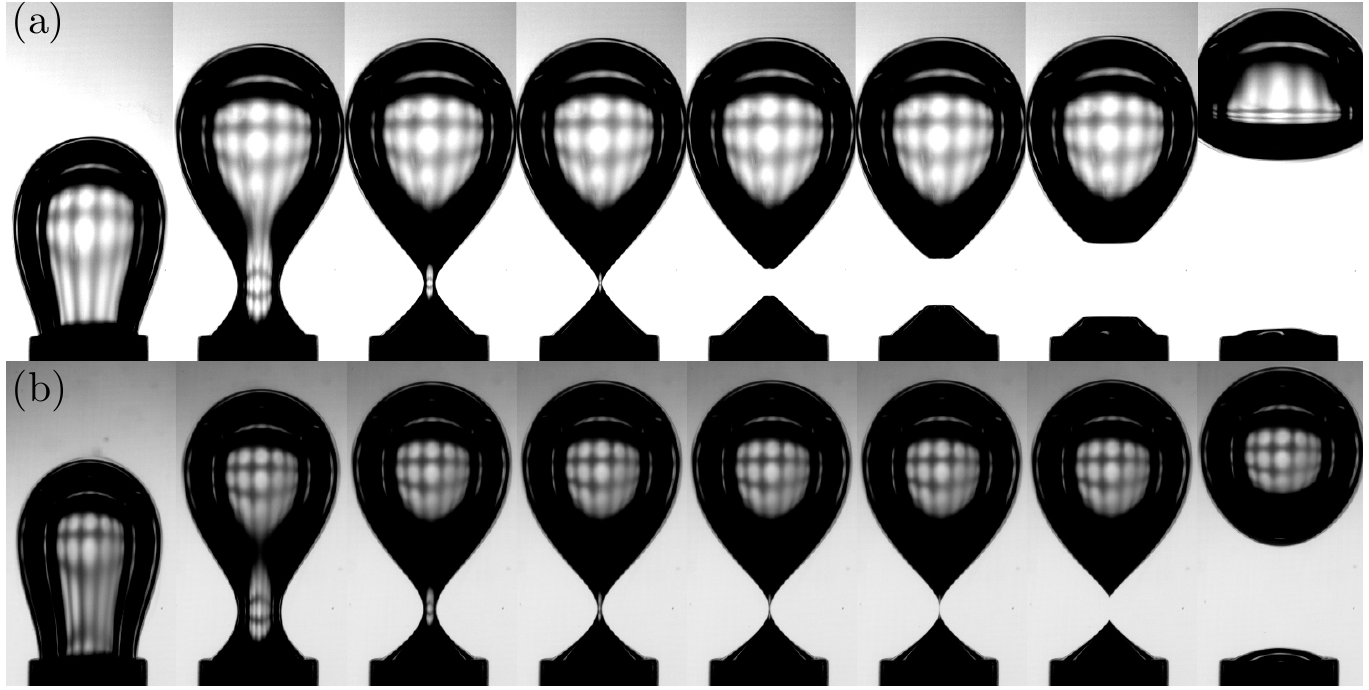}%
 \caption{(a) Thinning of a bubble in a 75/25 wt\% water/glycerol solvent of viscosity $\eta_{\rm s} = 2.14\,{\rm mPa\,s}$ and (b) in a 10/90 wt\% water/glycerol solvent of viscosity $\eta_{\rm s} = 213\,{\rm mPa\,s}$.}
 \label{fgr:SM_Fig4}
  \end{center}
 \end{figure}


\noindent {\Large {\textsc{Thinning in increasing concentrations of polymer solutions}}}

\medskip

We show in Fig. \ref{fgr:SM_Fig5} the thinning in polymer solutions when time $t > t_{\rm c}$. For low concentrations of 4000K PEO ($c \leq 0.02\%$), this time corresponds to when the bubble has already separated. However, for $c \geq 0.05\%$, the air filament is visible. The initial thickness of this filament increases with the polymer concentration, as illustrated in Fig. \ref{fgr:SM_Fig5}.

\medskip

 \begin{figure}[h!]
 \begin{center}
 \includegraphics[width=0.95\textwidth]{figures_sm/4000K PEO.pdf}%
 \caption{Profiles observed during the thinning of air bubbles in polymer solutions of 4000K PEO with mass concentrations from $c = 0\%$ to $c = 1\%$, prepared in a 75/25 wt\% water/glycerol solvent. The pictures are taken slightly after $t_{\rm c}$.}
 \label{fgr:SM_Fig5}
  \end{center}
 \end{figure}

\newpage

\noindent {\Large {\textsc{Thinning of a solution of 300K PEO}}}

\medskip

\begin{figure}[h!]
 \begin{center}
 \includegraphics[width=0.6\textwidth]{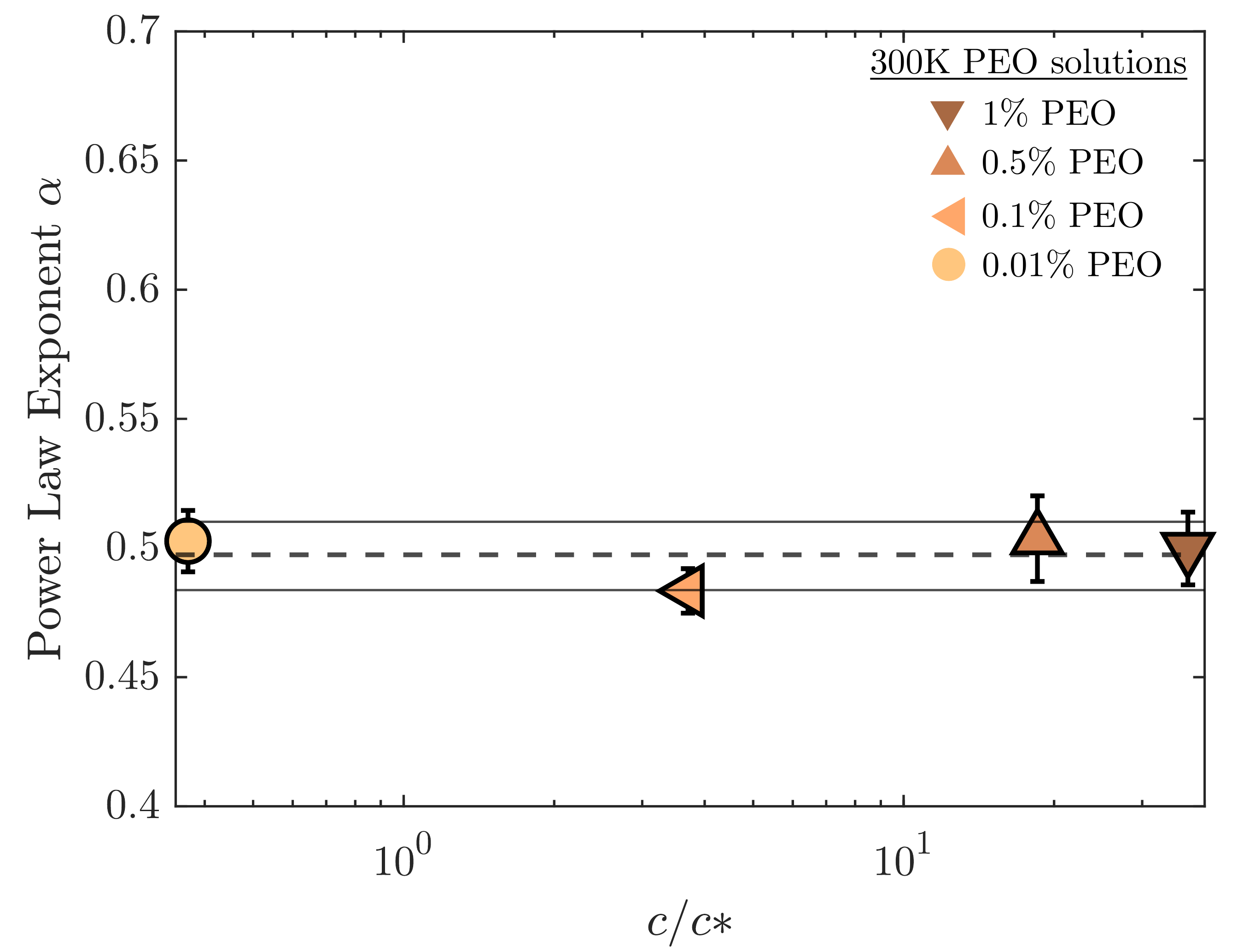}%
 \caption{Exponents $\alpha$ extracted from the power-law $h_{min} = A(t_{\rm c}-t)^\alpha$ for the experiments performed in the polymers solution of rescaled concentrations $c/c^* = 0.37$ to $c/c^* = 37$, corresponding to $c = 0.01\%$ to $c = 1\%$. The horizontal dashed line indicates the exponent $\alpha=0.497 \pm 0.013$. The yellow shaded region is the standard deviation of the measured exponents.}
 \label{fgr:SM_Fig6}
  \end{center}
 \end{figure}

To examine the influence of polymer molecular weight on the thinning dynamics, we prepare solutions of 300K polymer in a water/glycerol solvent. The thinning of the solution was then fitted with the power-law discussed in the main article, and is shown in Fig. \ref{fgr:SM_Fig6}. The exponents we recover are $0.497 \pm 0.013$, similar to the exponents we obtain for the solvent and the 4000K PEO solutions.

\newpage

\noindent {\Large {\textsc{Relaxation Modulus of the polymer solutions}}}

\medskip

\begin{figure}[h!]
 \begin{center}
 \includegraphics[width=0.6\textwidth]{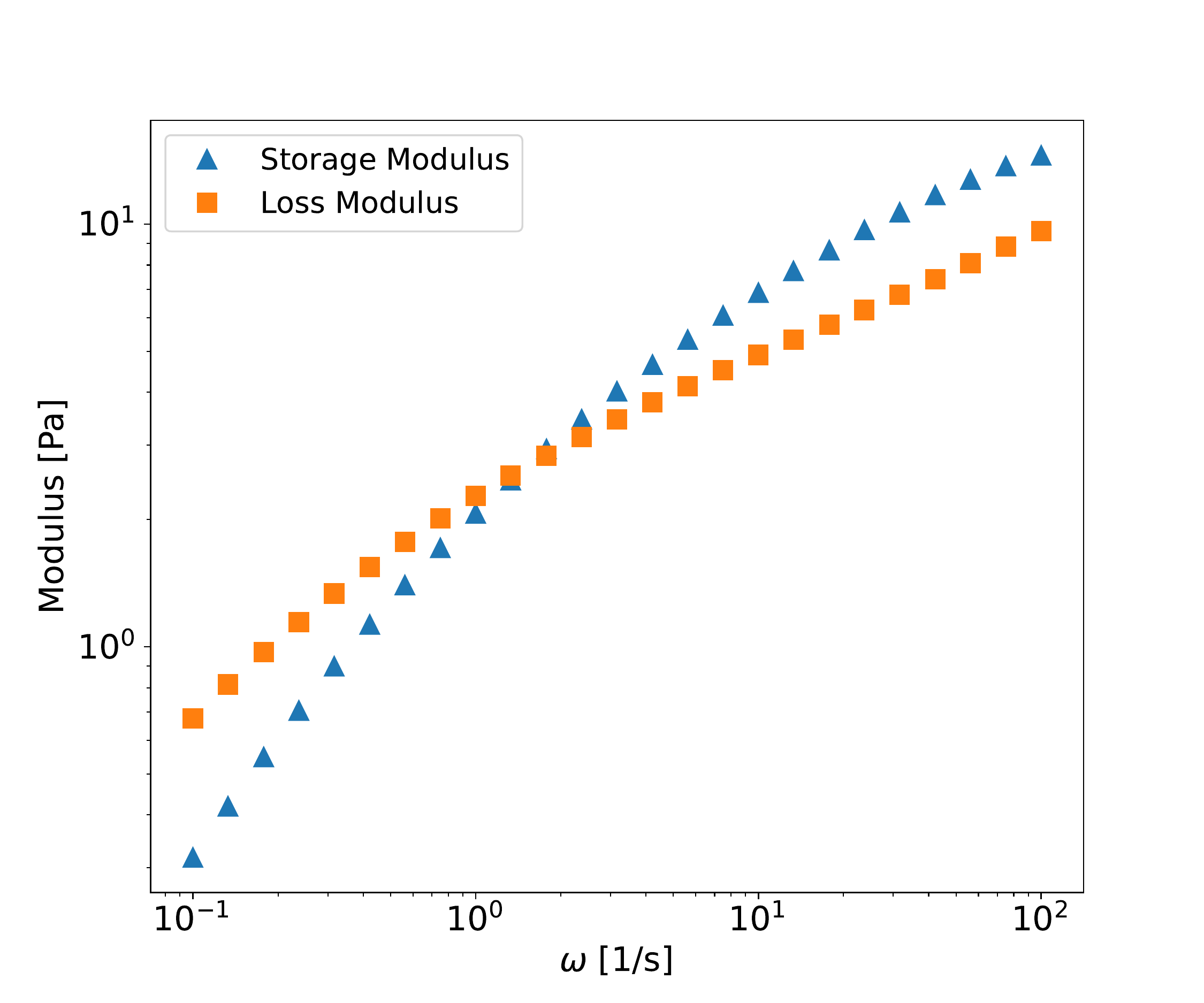}%
 \caption{The storage modulus, G' and the loss modulus, G``, measured for the 4000K polymer solution of concentration c = 1\%. The crossover time $1/\omega_{\rm c} \sim 0.75$ s is close to the relaxation time 0.5 s measured from the droplet pinch-off experiment.}
 \label{fgr:SM_RelaxationModulus}
  \end{center}
 \end{figure}

The relaxation modulus of a polymer solution is a reliable approach to measure the relaxation time of viscoelastic material at the crossover of the elastic modulus G' and viscous modulus G''. However, for the dilute and semi-dilute solutions used in this study ($c < 0.5 \%$ corresponding to $c/c* < 10$) the resulting G' and G'' signals are noisy. Because of the low viscosity, we are unable to see a crossover for solutions of $c \leq 0.5\%$. Similar observations were also noted for the relaxation modulus measurements of PEO solutions by Varma \textit{et al.}\cite{varma2022rheocoalescence} in a recent study. For the 4000K PEO of 1\% concentration, we measure the crossover frequency $\omega_{\rm c}$ of G' and G'', as shown in Fig. \ref{fgr:SM_RelaxationModulus}. The relaxation time is $\lambda_{\rm R} = 1/\omega_{\rm c}$. The time of crossover, $1/\omega_{\rm c} \sim 0.75$ s, is comparable to the relaxation time measured from droplet pinch-off. Hence, the droplet pinch-off offers a reliable alternate way to measure the relaxation time of the polymer solutions.

\noindent {\Large {\textsc{Movies of the experiments.}}}
\bigskip
\\
\noindent - Movie corresponding to pinch-off of a bubble in a 75/25 wt\% Water/Glycerol solvent at 10000 fps and 100000 fps.\\
\smallskip
\noindent - Movie corresponding to pinch-off of a bubble in a 0.01\% 4000K PEO in 75/25 wt\% Water/Glycerol Solvent at 10000 fps and 100000 fps.\\
\smallskip
\noindent - Movie corresponding to pinch-off of a bubble in a 1\% 4000K PEO in 75/25 wt\% Water/Glycerol Solvent at 10000 fps and 100000 fps.\\

\newpage

\bibliographystyle{plain}
\bibliography{bubble_pinchoff_sm}